\documentclass[english,preprint,showpacs,preprintnumbers,amsmath,amssymb,prb]{revtex4}
\usepackage[T1]{fontenc}
\usepackage[latin9]{inputenc}
\usepackage{graphicx}

\usepackage{dcolumn}
\usepackage{bm}
\usepackage{epsfig}\usepackage{subfigure}

\usepackage{babel}
\begin{document}

\title{Connection Between Minimum of Solubility and Temperature of Maximum
Density in an Associating Lattice Gas Model }

\author{Marcia M. Szortyka}

\affiliation{Departamento de F\'isica, Universidade Federal de Santa Catarina,
88010-970, Florian\'opolis, SC, Brazil}

\email{szortyka@gmail.com}

\author{Mauricio Girardi}

\affiliation{Universidade Federal de Santa Catarina. Rua Pedro Jo\~ao Pereira, 150
- Mato Alto, 88900-000, Ararangu\'a SC, Brazil}

\email{mauricio.girardi@ararangua.ufsc.br}

\author{Vera B. Henriques}

\affiliation{Instituto de F\'isica, Universidade de S\~ao Paulo, Caixa Postal 66318,
05315-970, S\~ao Paulo, SP, Brazil}

\email{vhenriques@if.usp.br}

\author{Marcia C. Barbosa}

\affiliation{Instituto de F\'isica, Universidade Federal do Rio Grande do Sul -
Caixa Postal 15051, CEP 91501-970, Porto Alegre, RS, Brazil}

\email{marcia.barbosa@ufrgs.br}

\begin{abstract}

In this paper we investigate the solubility of a hard - sphere gas in a solvent modeled as an associating lattice gas (ALG).
The solution phase diagram for solute at 5\% is compared with the phase diagram of the original solute free model. Model properties are investigated through Monte Carlo simulations and a cluster approximation. The model solubility 
is computed via simulations and shown to exhibit a minimum as a function of temperature. The line of minimum solubility (TmS) coincides with the 
line of maximum density (TMD) for 
different solvent chemical potentials.

\end{abstract}


\maketitle

\section{Introduction}

Solubility is the capacity of a given solute to form a homogeneous
solution in a given solvent. The solubility of one substance in another
is determined by the balance of intermolecular forces between the
solvent and solute, and the entropy change that accompanies the solvation.
Factors such as temperature and pressure will alter this balance,
thus changing the solubility.

In the process of dissolving a solute, heat is required to break the
intermolecular forces between the solute particles and heat is given
off by forming bonds between solute and solvent particles. Consequently,
increasing the temperature can make it easier or harder to dissolve
solute particles in a solvent.

In the case of dissolving solids, the energy required to break the
intermolecular forces between the solute particles is usually higher
than the energy liberated by forming bonds with the solvent, therefore,
the ability of the solvent to dissolve the solid solute increases
with the temperature. 

In the dissolution of gases in liquid solvents
the scenario is more complex.  For a low density gas phase, the 
solubility is the equilibrium constant of the process in
which a solute particle dissolved in the liquid evaporates
into the gas phase. This quantity can be express in 
terms of the inverse of the 
the Henry's constant. The equilibrium constant
can be computed by the balance between the 
solute-solute,  solvent-solvent and solute-solvent
energies, namely $W=(w_{solute-solute}+w_{solv-solv}-2w_{solute-solv})$, and
it is proportional to $e^{W/k_BT}$.
Therefore, if $W>0$, the solubility increases with temperature, otherwise it decreases. This 
approach, based in a simple "Bragg-Williams approximation", does
not allow for entropic effects besides the combinatory term that
differenciats solute from solvent particles~\cite{Hill}. 

This simple picture is valid for a number of solvents and solutes
but not for non-polar gases in water. The solubility of noble gases
in water decreases with  the temperature until a certain threshold,
and then it increases \cite{Iv05,Iv06,Fi97}. This experimental
result has also being observed in the solubility of hard spheres
simulated for effective \cite{Bu07, Buz03} and atomistic models of water
\cite{Ch08}.

Besides the unusual behavior of solubility, water also exhibits
other thermodynamic and dynamic anomalies. Unlike other liquids, water
does not contract upon cooling and the specific volume at ambient
pressure starts to increase when cooled below $T\approx4^{o}C$~\cite{Wa64,An76}.
Experimental data show that the diffusion constant $D$, increases
on compression at low temperatures $T$, up to a maximum, $D_{{\rm max}}(T)$
at $p=p_{D\mathrm{max}}(T)$ \cite{Wa64,An76} while for normal liquids
the diffusion coefficient decreases as the system is compressed. These
findings were also supported by simulations both in atomistic~\cite{Ne01,Er01}
and in effective models~\cite{Bu02,Rob96,Ol06a,Ol08b}.

Even though both experiments and simulations indicate that non-polar
gases exhibit an anomalous behavior in the solubility in the same
region of pressures and temperatures where other  anomalies are
present, a clear connection between these thermodynamic and dynamic
properties is still missing.

In this paper we aim to shade some light into this problem by computing
the solubility of non-interacting particles in a lattice model for
water. Even though there are a number of two and three dimensional lattice
models that would in principle exhibit the anomalies
present in water~\cite{Sa96,Rob96, Buz03,Al09,Fr07}, 
here water is represented by the associating lattice gas 
\cite{He05a,He05b,Ba07,Sz07,Gi07a,Sz09,Sz10a}
that have already shown the density and diffusion anomalies described
above. The solute is represented by pure hard sphere in an attempt
to model simple gases. Solubility versus temperature is calculated
and tested for the presence of a minimum. We show that for a hard
sphere solute, the temperatures of the maximum density and the temperatures
of the minimum solubility coincide. This result is explained in
the framework of the two length scales potentials and the presence
of thermodynamic and dynamic anomalies in water-like models.

The paper is organized as follows. In sec. II the model is presented,
in sec. III the results obtained by employing Monte Carlo simulations
are shown and in sec. IV the problem is analyzed by a Cluster approximation.
Our conclusions are summarized in sec. V.

\section{The Pure Water model - ALG}

\begin{figure}[h!]
\begin{centering}
\includegraphics[clip,scale=0.5]{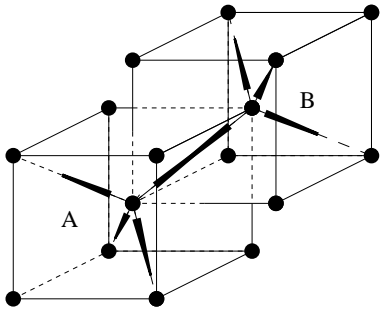} 
\par\end{centering}

\caption{The two possible configurations (A and B) of the water molecules
in a BCC lattice. \label{fig:lattice}}
\end{figure}

We consider the three dimensional Associating Lattice Gas Model (ALG) 
introduced by Girardi and coworkers \cite{Gi07a}. Dynamic and thermodynamic
properties of the system were previously obtained by Monte Carlo simulations\cite{Gi07a,Gi07,Sz10a}
and analytical methods \cite{Bu08}. The model consists of a body
centered cubic lattice where sites can be either occupied ($\sigma_{i}=1$)
by a water molecule or empty ($\sigma_{i}=0$). Besides the
occupational variables there are eight arm variables, $\tau_{i}$,
that describe the molecule orientation. Four
arms are the usual ice bonding arms, with $\tau_{i}=1$, while the other
four are inert arms, with $\tau_{i}=0$. The bonding and non-bonding arms
are distributed in a tetrahedral arrangement and there are two possible
configurations for a water particle, as shown in Fig. \ref{fig:lattice}.

The Hamiltonian that describes the system is given by
\begin{equation}
\mathcal{H}=\sum_{\langle i,j\rangle}\sigma_{i}\sigma_{j}\left(\epsilon+\gamma\,\tau_{i}\tau_{j}\right)\,\,;\label{E3d}
\end{equation}
 where $\sigma_{i}=0,1$ is the occupational variable, $\epsilon$
is a van der Waals like energy, $\gamma$ is the bond energy and
$\tau_{i}=0,1$ corresponds to the arm variable that represents the possibility
of a bond between two nearest-neighboring (NN) particles. A bond is
formed when two neighboring particles have bonding arms ($\tau=1$) pointing to each other. The parameters are chosen to be $\epsilon>0$
and $\gamma<0$, which implies an energetic penalty on neighbors
that do not form a bond.

\begin{figure}[h!]
\begin{centering}
\includegraphics[clip,scale=0.5]{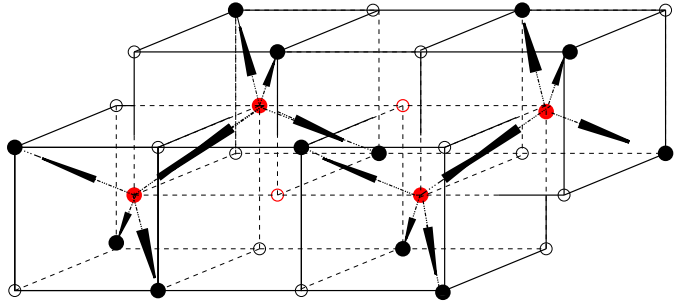} 
\par\end{centering}

\caption{In the ordered low density liquid phase half of the lattice is occupied
by particles. Water (filled circles) and empty sites (open circles)\label{fig:faseldl}}
\end{figure}

\begin{figure}[h!]
\begin{centering}
\includegraphics[clip,scale=0.5]{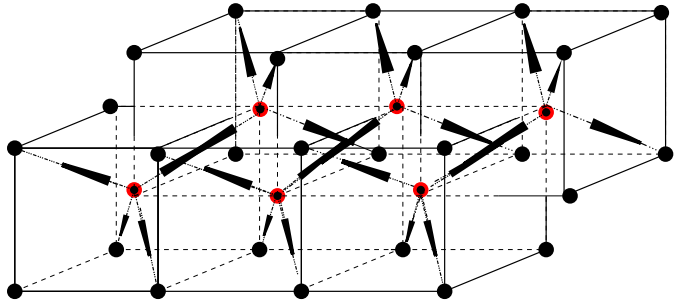} 
\par\end{centering}

\caption{In the ordered high density liquid phase the lattice is fully occupied
by particles. \label{fig:fasehdl}}
\end{figure}

The ground state of the pure solvent system can be inferred from 
inspection of the equation \ref{E3d}, and taking into account an external
chemical potential $\mu$. At zero temperature, the grand potential
per volume is $\Omega=e-\mu\rho$, where $\rho$ is water density
and $e=\mathcal{H}/V$. At very low values of the chemical potential,
the lattice is empty and the system is in the gas phase. As the chemical
potential increases, coexistence between a gas phase, with $\rho_{gas}=0$, and a low density liquid ($LDL$) with
$\rho_{LDL}=1/2$ is reached, at $\mu_{gas-LDL}=2(\varepsilon+\gamma)$. In the $LDL$
phase each molecule has only four neighboring molecules, forming four hydrogen bonds
(HBs), and the grand potential per volume is $\Omega_{LDL}=\varepsilon+\gamma-\mu/2$.
As the chemical potential increases even further a competition between
the chemical potential that favors filling up the lattice and the
HB penalty that favors molecules with just four $NN$ appears.
At $\mu_{LDL-HDL}=6\varepsilon+2\gamma$, the $LDL$ phase coexists
with a disordered high density liquid ($HDL$), with $\rho_{HDL}=1$. In the $HDL$ phase,
each molecule has eight $NN$ occupied sites, but forms only four
$HBs$. The other four non-bonded molecules contribute with positive energy, which can
be viewed as an effective weakening of the hydrogen bonds due to distortions
of the electronic orbitals of the bonded molecules. The grand potential
per volume is then $\Omega_{HDL}=4\varepsilon+2\gamma-\mu$. Both
phases are illustrated in Fig. \ref{fig:faseldl} and Fig. \ref{fig:fasehdl}. 

The non-zero temperature properties of the model were obtained from
Monte Carlo grand canonical simulations~\cite{Sz10a} in a previous publication, for $\gamma/\varepsilon=-2$.
Reduced parameters were defined as 
\begin{eqnarray}
\overline{T} & = & \frac{k_{B}\; T}{\varepsilon}\;\;,\nonumber \\
\overline{\mu} & = & \frac{\mu}{\varepsilon}\;\;.\label{par}
\end{eqnarray}

The chemical potential versus temperature phase diagram of
this water model is illustrated in the inset of Fig. \ref{fig:mu5}.
Continuous transitions were investigated through analysis of the specific
heat, of energy cumulants and of the order parameters. First-order transition
points were located from hysteresis of the system density as a function
of the chemical potential.

At low chemical potentials and low temperatures there is a first-order
transition (circles) between the gas and LDL phases. As the temperature
is increased this transition becomes continuous at a bicritical
point. At intermediate chemical potentials and high temperatures there
is a continuous transition (dashed line) between the gas and the
disordered phase. At high chemical potential and low temperatures
there is a first-order phase transition (squares) between the LDL
and the HDL phases. This transition becomes critical at a tricritical
point (higher temperatures). This system also presents the density
anomalous behavior observed in liquid water. At fixed pressure, the
density increases as the temperature is decreased, reaches a maximum
and decreases at lower temperatures. The line of the temperatures
of maximum density (triangles) is illustrated in the inset of
Fig. \ref{fig:mu5}.

\section{Non-polar solute in ALG solvent - simulations}

We investigate the solution phase diagram for inert solute particles. Non-polar solutes are modeled
as non-interacting hard spheres. Thus each lattice site can be empty
or occupied either by a water particle or by a non-polar solute. Simulations were run
as follows: solvent molecules are inserted or removed via a
Grand-Canonical Monte Carlo algorithm, while a fixed number of solute particles are allowed
to diffuse throughout the lattice, following the Metropolis prescription.

\begin{figure}[h!]
\begin{centering}
\includegraphics[clip,scale=0.4]{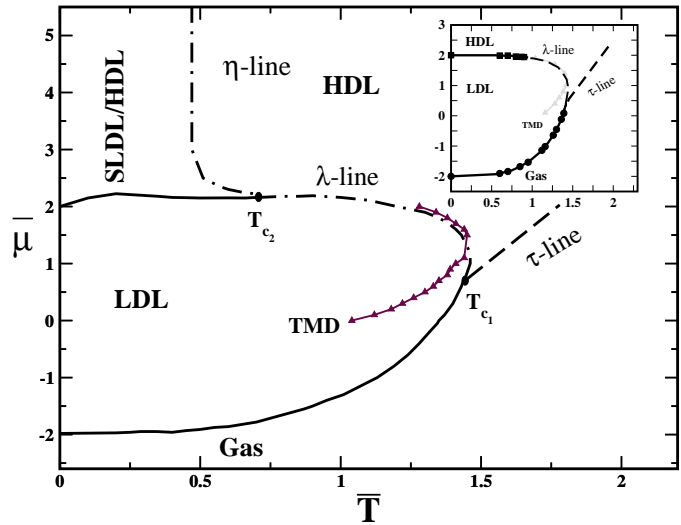} 
\par\end{centering}

\caption{Chemical potential vs. temperature phase diagram for the original
ALG model (inset) and ALG with 5$\%$ solute concentration.}

\label{fig:mu5} 
\end{figure}

\begin{figure}[h!]
\begin{centering}
\includegraphics[clip,scale=0.5]{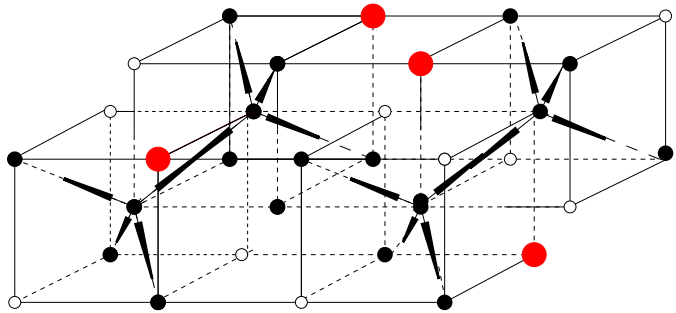} 
\par\end{centering}

\caption{In the ordered low density liquid phase half of the lattice is occupied
by particles and the other half contains solute.}

\label{fig:faseldl-solute} 
\end{figure}

Fig. \ref{fig:mu5} illustrates the effect on the water
chemical potential versus temperature phase diagram upon the introduction of $5\%$ solute ($\rho_{solute}=0.05$ in volume fraction). At $\overline {T} = 0$, the solute will be randomly distributed in the lattice.
In the gas phase, because solute is inert, the semi grand-potential
free energy per volume is the same as the grand potential of the solute free case, $\Omega_{gas}=0$.
In the LDL phase, the solute particles occupy the empty sites which correspond to half the lattice: for 5$\%$ of solute particles, 10$\%$ of the empty
sites are occupied, as illustrated in Fig. \ref{fig:faseldl-solute}.
Consequently, the density and the semi grand potential free energy per
volume are the same as that of pure water, $\rho_{LDL}=1/2$ and $\Omega_{LDL}=\varepsilon+\gamma-1/2$
respectively. The same occurs in the phase boundary between the
gas and the LDL phases at zero temperature, where $\mu_{gas-LDL}=2$.
This semi grand-potential free energy per volume is only correct if the
solute occupies up to 50$\%$ of the lattice. 

\begin{figure}[h!]
\begin{centering}
\includegraphics[clip,scale=0.5]{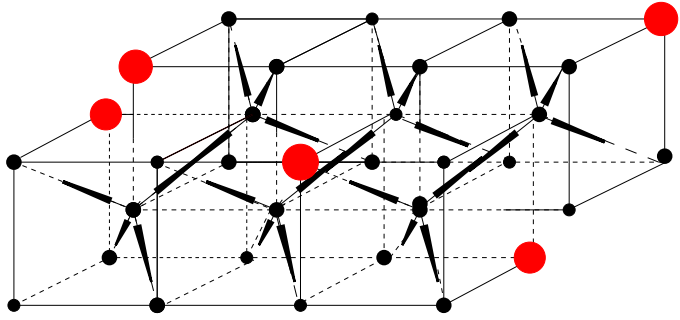} 
\par\end{centering}

\caption{In the ordered high density liquid phase the lattice is completely
occupied either by solute or water particles. Solutes are the bigger
circles. }

\label{fig:fasehdl-solute} 
\end{figure}

In the HDL phase the addition of the solute changes the density and
the grand potential per volume when compared with the free solute
case. At zero temperature, in the solute free case, the lattice is full.
As solute is added the particles replace solvent particles, changing
the energy, as can be seen in Fig. \ref{fig:fasehdl-solute}. The solute can occupy a number of different configurations
but it will choose the structure with the lowest grand potential free
energy. In order to identify which is this configuration we compare
two cases: all the solute forming a large cluster and the solute distributed
randomly but not  occupying adjacentes sites. In 
both cases
$\rho_{HDL}=1-\rho_{solute}$  but the 
grand potential free energy is
given  respectively by $\Omega_{HDL}^{cluster}=2(3\varepsilon+2\gamma)-(4\varepsilon+2\gamma)\rho_{solute}-\mu(1-\rho_{solute})$ and by $\Omega_{HDL}^{random}=4\varepsilon+2\gamma-(6\varepsilon+2\gamma)\rho_{solute}-\mu(1-\rho_{solute})$. The last case is the scenario with lower grand potential free
energy, therefore 
Even thought the density and the grand potential per volume is different
from the free solute case, the phase boundary between the LDL and
the HDL phases is given by the same value as that of pure water,
namely $\mu_{LDL-HDL}=6\varepsilon+2\gamma$.

At low temperature, and $\overline{T}\neq0$, the phase boundary between
the LDL and the gas phase depends on the entropy related to the addition
of the solute. In the gas phase the solute can be located
at any site, while in the LDL phase the solute goes only to the empty half-lattice (see Fig.~\ref{fig:faseldl-solute}), which makes
the entropy of the gas phase larger than the entropy of the LDL phase. This stabilizes the gas phase, moving up
the chemical potential of the gas-LDL phase boundary with
respect to the free solute case.

As to the LDL-HDL phase boundary at low $\overline{T}\neq0$, distorion of bonds in either phase is small, so that at very low temperatures no entropic
difference between the two phases due to the presence of the solute
is observed. However, as the temperature is increased and the bond network becomes more disordered, the entropy increment caused by solute is larger in the 
LDL phase, thus shifting the phase boundary to higher chemical potentials.

The critical lines illustrated in Fig. \ref{fig:mu5} were obtained
from the peak in the specific heat and through Binder cumulant methods \cite{Bi81, Landau}.
The critical line between the disordered phase
and the HDL phase, at high temperatures ($\tau$ line), and the critical
line between the HDL phase and the LDL phase ($\lambda$ line) are similar to the
lines obtained in the solute free case \cite{Sz10a}. The
shift in these lines to lower temperatures is due to the higher entropy
of the solute in the disordered phase.

Besides the $\lambda$'s and the $\tau$ critical lines present even
if no solute is present, the system exhibits a new critical line, which we call $\eta$,
at high chemical potentials. In order to inspect the new phase, we investigate the solute spatial distribution through its radial
distribution function, $g(r)$, given by 
\begin{eqnarray}
g(\bar{r})=\frac{\langle\sum_{i=1}^{N}\sum_{j>i}^{N}{\sigma_{i}\sigma_{j}\delta[\bar{r},(\bar{r}_{i}-\bar{r}_{j})]}\rangle_{\bar{T}}}{\langle\sum_{i=1}^{N}\sum_{j>i}^{N}{\sigma_{i}\sigma_{j}\delta[\bar{r},(\bar{r}_{i}-\bar{r}_{j})]}\rangle_{\infty}}\;.\label{gr}
\end{eqnarray}
 where the mean values denoted by $<...>_{\overline T}$ are obtained at constant
temperature for $T=\overline{T}$ and $\overline{T}\rightarrow\infty$, and 
$\delta$ is the Kronecker's delta.

\begin{figure}[h!]

\begin{centering}
\includegraphics[clip,scale=0.4]{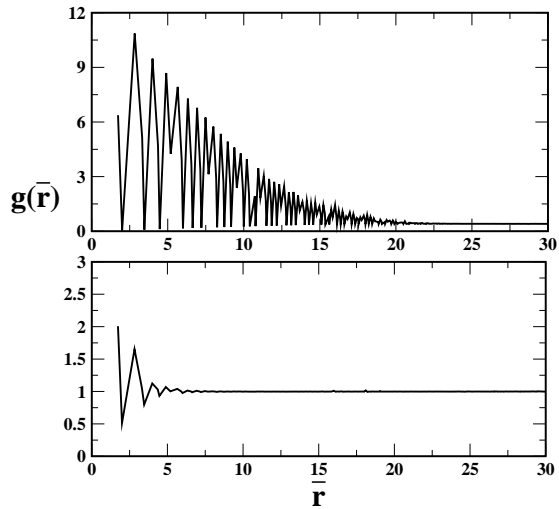} 
\par\end{centering}

\caption{Solute - solute radial distribution function, g(r), for solute concentration
5$\%$. From top to bottom $\bar{T}=0.40$ and $\bar{T}=0.70$, chemical
potential $\bar{\mu}=3.00$. At low temperature g(r) exhibits some
structural order while at high temperature g(r) is liquid like. }

\label{fig:gr5} 
\end{figure}

Figure \ref{fig:gr5} illustrates the radial distribution function
for 5$\%$ of solute for two different temperatures, $\bar{T}=0.40$ and $\bar{T}=0.70$.
At high temperatures the radial distribution is characteristic of a disordered, fluid-like system,
while at low temperatures solute presents an ordered (solid-like). 

\begin{figure}[h!]
\begin{centering}
\includegraphics[clip,scale=0.65]{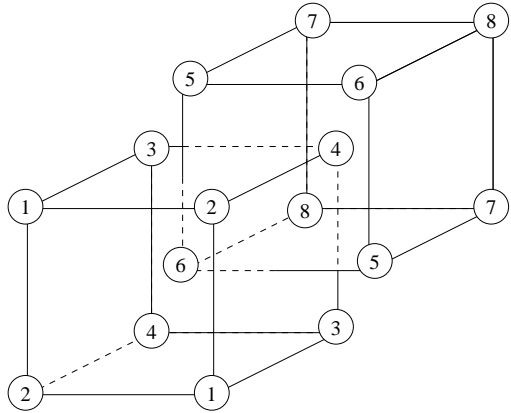} 
\par\end{centering}

\caption{The eight sub-lattices division. }

\centering{} \label{fig:subd} 
\end{figure}

\begin{figure}[h!]
\begin{centering}
\includegraphics[clip,scale=0.4]{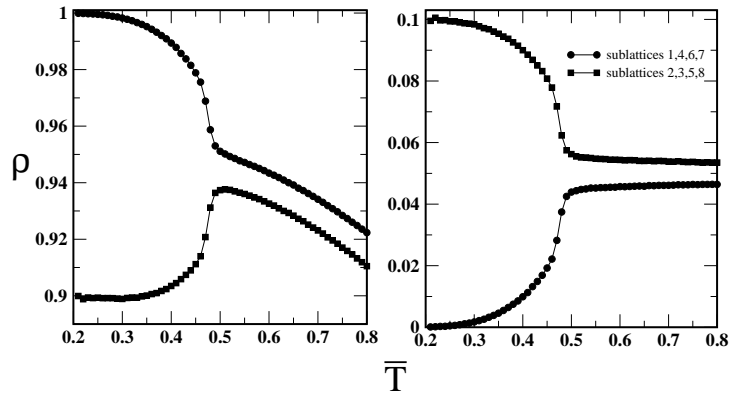} 
\par\end{centering}

\caption{Density of sub-lattice as a function of temperature for chemical
potential $\bar{\mu}=3.00$ and solute concentration 5$\%$. Left graph
shows solvent molecules and the right one, the solute. }

\label{fig:sub5} 
\end{figure}

To further analyse the structure of the new phase, we have measured sublattice solute and water densities. The lattice
is divided into eight sub-lattices, as shown in Fig. \ref{fig:subd}.
The result is shown in Fig. \ref{fig:sub5} for chemical potential fixed at
$\bar{\mu}=3.00$. At high temperatures, both solvent and solute are
randomly distributed, with no preferentially occupied sub-lattice.
As temperature decreases and the vertical critical line is crossed,
the solute particles become organized in four sub-lattices, while
the solvent fills the remaining space in the eight sublattices. Therefore,
the $\eta$-line separates the higher temperature homogeneous fluid phase from a dense
phase in which the solute organizes itself, named SLDL/HDL phase. 
It is not clear, however,
if inside each sub-lattice the solute particles are randomly distributed
or if they cluster, forming a high density solute and a low density
solute region. The behavior of the radial distribution function
suggests that the later is the case. This issue is checked through the cluster
approximation of the next section.

Besides the critical lines and the coexistence lines, this system
with and without solute, exhibits an anomalous behavior in the density.
As the temperature is decreased at constant pressure the density
increases, and at very low temperatures, it decreases, thus presenting
a maximum. The line of maximum density temperatures at different
chemical potentials is illustrated as the line
with triangles in Fig. \ref{fig:mu5}. Due to the entropic effects
the TMD line is located at higher temperatures when compared with
the solute free case.

Having explored the effect of introducing inert solute upon the solvent
chemical potential versus temperature phase diagram, we investigate
the solubility of these non-interacting particles. The solubility
is derived from the solute excess chemical potential which is computed
through Widom's insertion method namely 
\begin{eqnarray}
\mu_{ex}=-\bar{T}\ln\langle e^{-\beta e_{solute}}\rangle_{0}\;\; & .\label{wid}
\end{eqnarray}
where $e_{solute}$ represents the energy increment due to the added solute. Note that the solutes interact
only via excluded volume.
In the limit of low solute concentration, the solubility is then
given by
\begin{eqnarray}
\Gamma=exp\left(\frac{-\mu_{ex}}{\bar{T}}\right)\;\;.\label{sol}
\end{eqnarray}

\begin{figure}[h]
\begin{centering}
\includegraphics[clip,scale=0.45]{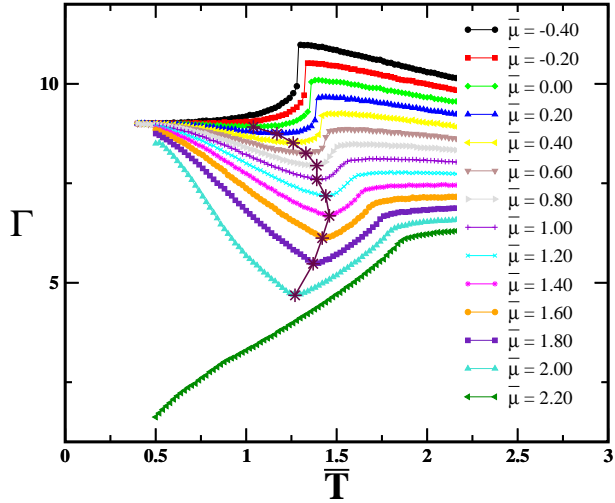} 
\par\end{centering}

\caption{Solute excess chemical potential vs. temperature for different values
of the water chemical potential. }

\label{fig:muex5} 
\end{figure}

Figure \ref{fig:muex5} illustrates the solubility parameter $\Gamma$
as a function of temperature for solute concentrations 5$\%$
for various chemical potentials. It can be seen that the solubility exhibits a minimum for a certain range of
chemical potentials. The temperatures
of minimum solubility (TmS) are shown with star symbols. The TMD and the TmS lines coincide. This
coincidence can be understood as follows.

The solubility is related with the amount of energy required to include
a solute particle in the system. In the gas phase, as the temperature
is increased, the solubility decreases because the density of the gas
phase increases with temperature making more difficult to include
an extra solute particle. In the HDL phase the density of solvent
decreases with the increase of temperature, therefore the solubility decreases
with the decrease of temperature. In the LDL phase a very interesting
behavior is observed. The density, which at zero temperature is $1/2$,
increases with the temperature. In this low temperature region the
increase of the solvent density leaves less space for including a new solute
particle and the solubility decreases. However, as the temperature
is increased even further, the density of the LDL phase reaches
a maximum and decreases due to entropic effects. Consequently, the
solubility reaches a minimum at the maximum of solvent density, and increases with
$\overline{T}$ as the system becomes less dense. The decrease of 
the solubility with the increase of the temperature
is also observed in a  minimal lattice models proposed by Widom and
collaborators~\cite{Ko99}.  The focus of these different studies on the model properties  have been on the relation between strength and range of the solvent mediated attraction, thus density effects, which would promote increasing solubility at larger temperatures, were left aside~\cite{Wi03,Ba10}.

Obviously the solubility governed solely by density effects is only possible
because the solute particles, in our study, are non interacting.

\section{Non-polar solute in ALG solvent - cluster variational approximation}

In this section, we present results obtained from the Cluster Variational Approximation, introduced
for the pure solvent version of this model by Buzano \emph{et al.
}\cite{Bu08}. The approximation is based on the assumption that an elementary tetrahedron
cluster of sites, connecting the four sub-lattices (Fig. \ref{fig:subd})
is thermodynamically representative.

The energy per site is written as
\[
w=\sum_{i,j,k,l}p_{ijkl}\mathcal{H}_{ijkl}\,,
\]
 where $p_{ijkl}$ is the joint probability for the configurations
of the cluster in the four sub-lattices $1$, $2$, $3$ and $4$,
and $\mathcal{H}_{ijkl}$ represents the Hamiltonian for the cluster.

For interactions restricted to nearest-neighbors the
Hamiltonian $\mathcal{H}_{ijkl}$ can be decomposed in the form 
\[
\mathcal{H}_{ijkl}=\mathcal{H}_{ij}+\mathcal{H}_{jk}+\mathcal{H}_{kl}+\mathcal{H}_{li}\,,
\]
 where the pair Hamiltonian can be written as 
\[
\mathcal{H}_{ij}=\epsilon\sigma_{i}\sigma_{j}-\gamma\sigma_{i}\sigma_{j}h_{ij}-\frac{\mu\sigma_{i}}{4}-\frac{\nu b_{i}}{4}\,.
\]

Here, $\sigma_{i}=1$ if site $i$ is occupied by a water molecule and $\sigma_{i}=0$, otherwise. The new state variable $b$ stands for solute occupation, with $b_{i}=1$ if site $i$ is occupied by a solute particle and $b_{i}=0$ otherwise.
$h_{ij}$ represents bond states and is equal to $1$ if there is a bond between
the two waters at sites $i$ and $j$, and equal to zero for non-bonding
waters. To the parameters of the previous section, hydrogen-bond interaction $\gamma$, non-bonding penalty $\epsilon>0$ and solvent chemical potential $\mu$ we add the solute chemical potential $\nu$. 

By minimizing the grand-canonical free energy 
\[
\frac{\omega}{T}=\sum_{i,j,k,l}p_{ijkl}\left[\frac{\mathcal{H}_{ijkl}}{T}+\ln p_{ijkl}-\frac{3}{4}\ln(p_{i}^{1}p_{j}^{2}p_{k}^{3}p_{l}^{4})\right]\,\,,
\]
 where $p_{i}^{1}=\sum_{jkl}p_{ijkl}$, $p_{j}^{2}=\sum_{ikl}p_{ijkl}$,
$p_{k}^{3}=\sum_{ijl}p_{ijkl}$ and $p_{l}^{4}=\sum_{ijk}p_{ijkl}$.
According to the natural iteration method \cite{kiku}, for fixed
values of $T,\,\epsilon,\,\gamma,\,\mu,$ and $\nu$, one may different quantities of interest, such as solvent and solute concentrations,
hydrogen bond fraction or sub-lattice densities.

In order to compare the cluster approximation results to previous simulation results, we search for the desired solute concentration ($\rho_{solute}$) by varying the corresponding solute chemical potential $\nu$. In certain ranges of the thermodynamic parameters, 
hysteresis loops were found.

\begin{figure}
\begin{centering}
\includegraphics[bb=20bp 20bp 400bp 400bp,clip,scale=0.45]{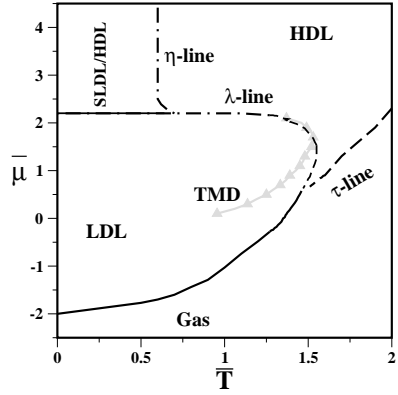} 
\par\end{centering}

\caption{\label{fig:13} Solvent reduced chemical potential $\mu$ versus reduced temperature
phase diagram. The Gas-LDL and LDL - SLDL/HDL coexistence phases,
critical and TMD lines are shown.}
\end{figure}

The phase diagram for $5\%$ solute concentration ($\rho_{solute}=0.05$)
is displayed in Fig. \ref{fig:13}. It is qualitatively similar to that
of simulations, and the new phase SLDL/HDL in which solute is ordered is present. In that phase, a solute droplet occupying a sub-lattice
forces the neighboring solvent to order as a LDL. All other non solvent molecules are in the HDL phase. In other words, as the
temperature is lowered and one crosses the fluid - SLDL/HDL line
for a given value of $\mu$, a second-order phase transition
takes place, and the solute, which was dispersed in the fluid phase,
begins the formation of sub-lattice clusters, intercalated with LDL water. Thus, the SLDL/HDL phase is a region of coexistence
between clustered and dispersed solute, or between LDL+solute and HDL-no solute.

\begin{figure}
\begin{centering}
\includegraphics[bb=0bp 0bp 380bp 380bp,clip,scale=0.45]{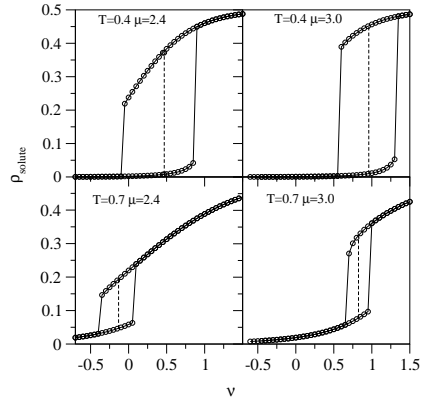} 
\par\end{centering}

\caption{\label{fig:14}Solute concentration $\rho_{solute}$ as a function
of the chemical potential $\nu$ for some values of temperature and
water chemical potential $\mu$ near the fluid - SLDL/HDL transition line.
Dashed lines give the coexistence lines.}
\end{figure}

The picture we propose for the SLDL/HDL phase is supported by behavior shown in Fig. \ref{fig:14}, for solute density as a function
of solute chemical potential $\nu$, at fixed $T$ and solvent chemical potential $\mu$, in the vicinity of
the HDL - SLDL/HDL line. As $\nu$ increases, the solute density
goes through a discontinuity. The diagrams show coexistence between a low 
density, with solute dispersed in HDL solvent, and a
high density, with solute grouped in LDL solvent.
Maxwell construction for the curves of Fig. \ref{fig:14},
yield the densities of the two coexisting phases shown in
Table 1. It can be seen that for $T=0.4$, for both values of the solvent chemical potential, as well as for $T=0.7$ and $\mu=2.4$, 
the case of $5\%$ solute density ($\rho_{solute}=0.05$) is between the low and high density limits ($\rho_l$ and $\rho_h$). The three cases correspond to points on the left of the fluid - SLDL/HDL $\eta$-line, although very near to the line in the case of $T=0.7$.
On the other hand, for $\mu=3.0$ and $T=0.7$ (clearly on the right
of the HDL -  SLDL/HDL $\eta$-line), the low density at coexistence is $\rho_{low}=0.078$, larger than $0.05$ implying
solute must be homogeneously distributed in the solution. 

\begin{table}
\begin{centering}
\includegraphics[bb=50bp 200bp 600bp 800bp,clip,scale=0.25]{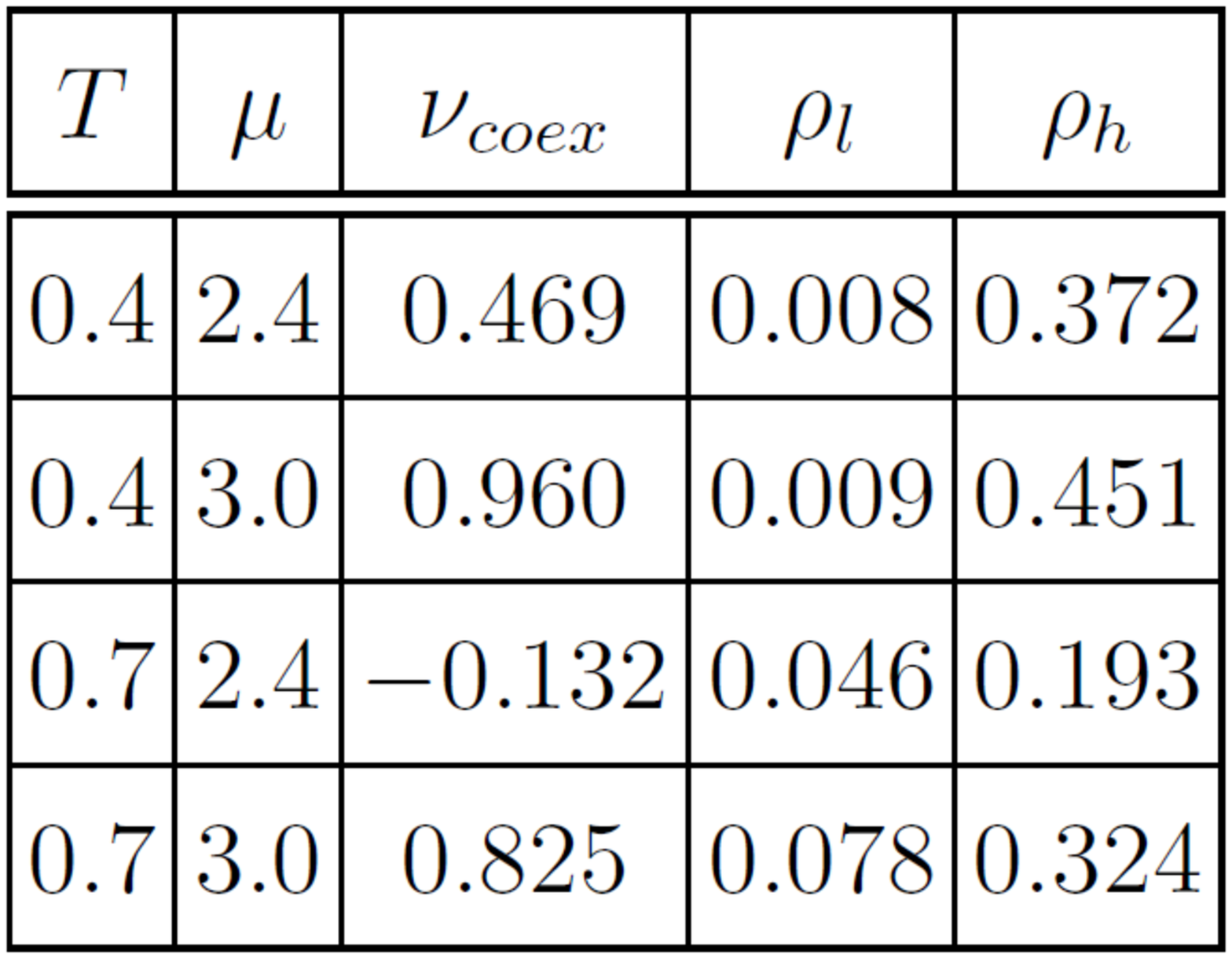}
\par\end{centering}

\caption{Solute chemical potential at the coexistence $\nu_{coex}$ and the
respective concentrations of solute in both phases.}
\end{table}

\section{Conclusions}

In this paper we have analyzed the effect of adding non interacting solute particles to the Associating Lattice Gas Model for water. We have shown
that the solute particles change the chemical potential versus
temperature phase diagram substantially. In the first place, by enhancing entropic
effects of the disordered and gas phases, addition of solute displaces 
the coexistence and critical lines accordingly.
Moreover, a new phase appears, with coexistence between HDL and LDL water induced by the presence of solute. Thus, in addition to the $\lambda$ and $\tau$ critical lines present in
the solute free case, a third order-disorder continuous $\eta$
transition at which the solute orders itself is present.

The solubility of this non interacting system was also computed and
we have found that, as the temperature is decreased for fixed solvent
chemical potential, the model solution exhibits a minimum in the solubility.
The temperature of minimum of solubility coincides with the temperature of maximum density. This coincidence
is due to the fact that the solute is purely hardcore. For hydrophobic
or hydrophilic solutes this coincidence should to be lost.

Our results suggest a very simple mechanism for understanding the
minimum of solubility in simple gases.

\subsection*{Acknowledgments}

This work is partially supported by CNPq, Capes, INCT-FCx and Universidade Federal de Santa Catarina.

\vspace{1cm}

\bibliographystyle{apsrev}

\begin{thebibliography}{33}
\expandafter\ifx\csname natexlab\endcsname\relax\def\natexlab#1{#1}\fi
\expandafter\ifx\csname bibnamefont\endcsname\relax
  \def\bibnamefont#1{#1}\fi
\expandafter\ifx\csname bibfnamefont\endcsname\relax
  \def\bibfnamefont#1{#1}\fi
\expandafter\ifx\csname citenamefont\endcsname\relax
  \def\citenamefont#1{#1}\fi
\expandafter\ifx\csname url\endcsname\relax
  \def\url#1{\texttt{#1}}\fi
\expandafter\ifx\csname urlprefix\endcsname\relax\def\urlprefix{URL }\fi
\providecommand{\bibinfo}[2]{#2}
\providecommand{\eprint}[2][]{\url{#2}}

\bibitem[{\citenamefont{Hill}(1960)}]{Hill}
\bibinfo{author}{\bibfnamefont{T.~L.} \bibnamefont{Hill}},
  \emph{\bibinfo{title}{Statistical Thermodynamics}}
  (\bibinfo{publisher}{Addison-Wesley Pub.}, \bibinfo{address}{New York},
  \bibinfo{year}{1960}), \bibinfo{edition}{1st} ed.

\bibitem[{\citenamefont{Ivanov et~al.}(2005)\citenamefont{Ivanov, Lebedeva,
  Abrosimov, and Ivanova}}]{Iv05}
\bibinfo{author}{\bibfnamefont{E.~V.} \bibnamefont{Ivanov}},
  \bibinfo{author}{\bibfnamefont{E.~J.} \bibnamefont{Lebedeva}},
  \bibinfo{author}{\bibfnamefont{V.~K.} \bibnamefont{Abrosimov}},
  \bibnamefont{and} \bibinfo{author}{\bibfnamefont{N.~G.}
  \bibnamefont{Ivanova}}, \bibinfo{journal}{J. Struct. Chem.}
  \textbf{\bibinfo{volume}{46}}, \bibinfo{pages}{253} (\bibinfo{year}{2005}).

\bibitem[{\citenamefont{Ivanov and Abrosimov}(2006)}]{Iv06}
\bibinfo{author}{\bibfnamefont{E.~V.} \bibnamefont{Ivanov}} \bibnamefont{and}
  \bibinfo{author}{\bibfnamefont{V.~K.} \bibnamefont{Abrosimov}},
  \bibinfo{journal}{Radiochemistry} \textbf{\bibinfo{volume}{48}},
  \bibinfo{pages}{244} (\bibinfo{year}{2006}).

\bibitem[{\citenamefont{Filipponi et~al.}(1997)\citenamefont{Filipponi, Bowron,
  Lobban, and Finney}}]{Fi97}
\bibinfo{author}{\bibfnamefont{A.}~\bibnamefont{Filipponi}},
  \bibinfo{author}{\bibfnamefont{D.~T.} \bibnamefont{Bowron}},
  \bibinfo{author}{\bibfnamefont{C.}~\bibnamefont{Lobban}}, \bibnamefont{and}
  \bibinfo{author}{\bibfnamefont{J.~L.} \bibnamefont{Finney}},
  \bibinfo{journal}{Phys. Rev. Lett.} \textbf{\bibinfo{volume}{79}},
  \bibinfo{pages}{1293} (\bibinfo{year}{1997}).

\bibitem[{\citenamefont{Buldyrev et~al.}(2007)\citenamefont{Buldyrev, Kumar,
  Debenedetti, Rossky, and Stanley}}]{Bu07}
\bibinfo{author}{\bibfnamefont{S.~V.} \bibnamefont{Buldyrev}},
  \bibinfo{author}{\bibfnamefont{P.}~\bibnamefont{Kumar}},
  \bibinfo{author}{\bibfnamefont{P.~G.} \bibnamefont{Debenedetti}},
  \bibinfo{author}{\bibfnamefont{P.~J.} \bibnamefont{Rossky}},
  \bibnamefont{and} \bibinfo{author}{\bibfnamefont{H.~E.}
  \bibnamefont{Stanley}}, \bibinfo{journal}{Proc. Natl. Acad. Sci. U.S.A.}
  \textbf{\bibinfo{volume}{104}}, \bibinfo{pages}{20177}
  (\bibinfo{year}{2007}).

\bibitem[{\citenamefont{Buzano and Pretti}(2003)}]{Buz03}
\bibinfo{author}{\bibfnamefont{C.}~\bibnamefont{Buzano}} \bibnamefont{and}
  \bibinfo{author}{\bibfnamefont{M.}~\bibnamefont{Pretti}},
  \bibinfo{journal}{J. Chem. Phys.} \textbf{\bibinfo{volume}{119}},
  \bibinfo{pages}{3791} (\bibinfo{year}{2003}).

\bibitem[{\citenamefont{Chatterjee et~al.}(2008)\citenamefont{Chatterjee,
  Debenedetti, Stillinger, and Lynden-Bell}}]{Ch08}
\bibinfo{author}{\bibfnamefont{S.}~\bibnamefont{Chatterjee}},
  \bibinfo{author}{\bibfnamefont{P.~G.} \bibnamefont{Debenedetti}},
  \bibinfo{author}{\bibfnamefont{F.~H.} \bibnamefont{Stillinger}},
  \bibnamefont{and} \bibinfo{author}{\bibfnamefont{R.~M.}
  \bibnamefont{Lynden-Bell}}, \bibinfo{journal}{J. Chem. Phys.}
  \textbf{\bibinfo{volume}{128}}, \bibinfo{pages}{124511}
  (\bibinfo{year}{2008}).

\bibitem[{\citenamefont{Waler}(1964)}]{Wa64}
\bibinfo{author}{\bibfnamefont{R.}~\bibnamefont{Waler}},
  \emph{\bibinfo{title}{Essays of natural experiments}}
  (\bibinfo{publisher}{Johnson Reprint}, \bibinfo{address}{New York},
  \bibinfo{year}{1964}).

\bibitem[{\citenamefont{Angell et~al.}(1976)\citenamefont{Angell, Finch, and
  Bach}}]{An76}
\bibinfo{author}{\bibfnamefont{C.~A.} \bibnamefont{Angell}},
  \bibinfo{author}{\bibfnamefont{E.~D.} \bibnamefont{Finch}}, \bibnamefont{and}
  \bibinfo{author}{\bibfnamefont{P.}~\bibnamefont{Bach}}, \bibinfo{journal}{J.
  Chem. Phys.} \textbf{\bibinfo{volume}{65}}, \bibinfo{pages}{3063}
  (\bibinfo{year}{1976}).

\bibitem[{\citenamefont{Netz et~al.}(2001)\citenamefont{Netz, Starr, Stanley,
  and Barbosa}}]{Ne01}
\bibinfo{author}{\bibfnamefont{P.~A.} \bibnamefont{Netz}},
  \bibinfo{author}{\bibfnamefont{F.~W.} \bibnamefont{Starr}},
  \bibinfo{author}{\bibfnamefont{H.~E.} \bibnamefont{Stanley}},
  \bibnamefont{and} \bibinfo{author}{\bibfnamefont{M.~C.}
  \bibnamefont{Barbosa}}, \bibinfo{journal}{J. Chem. Phys.}
  \textbf{\bibinfo{volume}{115}}, \bibinfo{pages}{344} (\bibinfo{year}{2001}).

\bibitem[{\citenamefont{Errington and Debenedetti}(2001)}]{Er01}
\bibinfo{author}{\bibfnamefont{J.~R.} \bibnamefont{Errington}}
  \bibnamefont{and} \bibinfo{author}{\bibfnamefont{P.~G.}
  \bibnamefont{Debenedetti}}, \bibinfo{journal}{Nature (London)}
  \textbf{\bibinfo{volume}{409}}, \bibinfo{pages}{318} (\bibinfo{year}{2001}).

\bibitem[{\citenamefont{Buldyrev et~al.}(2002)\citenamefont{Buldyrev, Franzese,
  Giovambattista, Malescio, Sadr-Lahijany, Scala, Skibinsky, and
  Stanley}}]{Bu02}
\bibinfo{author}{\bibfnamefont{S.~V.} \bibnamefont{Buldyrev}},
  \bibinfo{author}{\bibfnamefont{G.}~\bibnamefont{Franzese}},
  \bibinfo{author}{\bibfnamefont{N.}~\bibnamefont{Giovambattista}},
  \bibinfo{author}{\bibfnamefont{G.}~\bibnamefont{Malescio}},
  \bibinfo{author}{\bibfnamefont{M.~R.} \bibnamefont{Sadr-Lahijany}},
  \bibinfo{author}{\bibfnamefont{A.}~\bibnamefont{Scala}},
  \bibinfo{author}{\bibfnamefont{A.}~\bibnamefont{Skibinsky}},
  \bibnamefont{and} \bibinfo{author}{\bibfnamefont{H.~E.}
  \bibnamefont{Stanley}}, \bibinfo{journal}{Physica A}
  \textbf{\bibinfo{volume}{304}}, \bibinfo{pages}{23} (\bibinfo{year}{2002}).

\bibitem[{\citenamefont{Roberts and Debenedetti}(1996)}]{Rob96}
\bibinfo{author}{\bibfnamefont{C.~J.} \bibnamefont{Roberts}} \bibnamefont{and}
  \bibinfo{author}{\bibfnamefont{P.~G.} \bibnamefont{Debenedetti}},
  \bibinfo{journal}{J. Chem. Phys.} \textbf{\bibinfo{volume}{105}},
  \bibinfo{pages}{658} (\bibinfo{year}{1996}).

\bibitem[{\citenamefont{de~Oliveira et~al.}(2006)\citenamefont{de~Oliveira,
  Netz, Colla, and Barbosa}}]{Ol06a}
\bibinfo{author}{\bibfnamefont{A.~B.} \bibnamefont{de~Oliveira}},
  \bibinfo{author}{\bibfnamefont{P.~A.} \bibnamefont{Netz}},
  \bibinfo{author}{\bibfnamefont{T.}~\bibnamefont{Colla}}, \bibnamefont{and}
  \bibinfo{author}{\bibfnamefont{M.~C.} \bibnamefont{Barbosa}},
  \bibinfo{journal}{J. Chem. Phys.} \textbf{\bibinfo{volume}{124}},
  \bibinfo{pages}{084505} (\bibinfo{year}{2006}).

\bibitem[{\citenamefont{de~Oliveira et~al.}(2008)\citenamefont{de~Oliveira,
  Franzese, Netz, and Barbosa}}]{Ol08b}
\bibinfo{author}{\bibfnamefont{A.~B.} \bibnamefont{de~Oliveira}},
  \bibinfo{author}{\bibfnamefont{G.}~\bibnamefont{Franzese}},
  \bibinfo{author}{\bibfnamefont{P.~A.} \bibnamefont{Netz}}, \bibnamefont{and}
  \bibinfo{author}{\bibfnamefont{M.~C.} \bibnamefont{Barbosa}},
  \bibinfo{journal}{J. Chem. Phys.} \textbf{\bibinfo{volume}{128}},
  \bibinfo{pages}{064901} (\bibinfo{year}{2008}).

\bibitem[{\citenamefont{Sastry et~al.}(1996)\citenamefont{Sastry, Debenedetti,
  and Sciortino}}]{Sa96}
\bibinfo{author}{\bibfnamefont{S.}~\bibnamefont{Sastry}},
  \bibinfo{author}{\bibfnamefont{P.~G.} \bibnamefont{Debenedetti}},
  \bibnamefont{and}
  \bibinfo{author}{\bibfnamefont{F.}~\bibnamefont{Sciortino}},
  \bibinfo{journal}{Phys. Rev. E} \textbf{\bibinfo{volume}{53}},
  \bibinfo{pages}{6144} (\bibinfo{year}{1996}).

\bibitem[{\citenamefont{Almarza et~al.}(2009)\citenamefont{Almarza, Capitan,
  Cuesta, and Lomba}}]{Al09}
\bibinfo{author}{\bibfnamefont{N.~G.} \bibnamefont{Almarza}},
  \bibinfo{author}{\bibfnamefont{J.~A.} \bibnamefont{Capitan}},
  \bibinfo{author}{\bibfnamefont{J.~A.} \bibnamefont{Cuesta}},
  \bibnamefont{and} \bibinfo{author}{\bibfnamefont{E.}~\bibnamefont{Lomba}},
  \bibinfo{journal}{J. Chem. Phys} \textbf{\bibinfo{volume}{131}},
  \bibinfo{pages}{124506} (\bibinfo{year}{2009}).

\bibitem[{\citenamefont{Franzese and Stanley}(2007)}]{Fr07}
\bibinfo{author}{\bibfnamefont{G.}~\bibnamefont{Franzese}} \bibnamefont{and}
  \bibinfo{author}{\bibfnamefont{H.~E.} \bibnamefont{Stanley}},
  \bibinfo{journal}{J. Phys.: Condens. Matter} \textbf{\bibinfo{volume}{19}},
  \bibinfo{pages}{205126} (\bibinfo{year}{2007}).

\bibitem[{\citenamefont{Henriques and Barbosa}(2005)}]{He05a}
\bibinfo{author}{\bibfnamefont{V.~B.} \bibnamefont{Henriques}}
  \bibnamefont{and} \bibinfo{author}{\bibfnamefont{M.~C.}
  \bibnamefont{Barbosa}}, \bibinfo{journal}{Phys. Rev. E}
  \textbf{\bibinfo{volume}{71}}, \bibinfo{pages}{031504}
  (\bibinfo{year}{2005}).

\bibitem[{\citenamefont{Henriques et~al.}(2005)\citenamefont{Henriques,
  Guissoni, Barbosa, Thielo, and Barbosa}}]{He05b}
\bibinfo{author}{\bibfnamefont{V.~B.} \bibnamefont{Henriques}},
  \bibinfo{author}{\bibfnamefont{N.}~\bibnamefont{Guissoni}},
  \bibinfo{author}{\bibfnamefont{M.~A.} \bibnamefont{Barbosa}},
  \bibinfo{author}{\bibfnamefont{M.}~\bibnamefont{Thielo}}, \bibnamefont{and}
  \bibinfo{author}{\bibfnamefont{M.~C.} \bibnamefont{Barbosa}},
  \bibinfo{journal}{Mol. Phys.} \textbf{\bibinfo{volume}{103}},
  \bibinfo{pages}{3001} (\bibinfo{year}{2005}).

\bibitem[{\citenamefont{Balladares et~al.}(2009)\citenamefont{Balladares, M.,
  Henriques, and Barbosa}}]{Ba07}
\bibinfo{author}{\bibfnamefont{A.~L.} \bibnamefont{Balladares}},
  \bibinfo{author}{\bibnamefont{M.}},
  \bibinfo{author}{\bibfnamefont{V.}~\bibnamefont{Henriques}},
  \bibnamefont{and} \bibinfo{author}{\bibfnamefont{M.~C.}
  \bibnamefont{Barbosa}}, \bibinfo{journal}{J. Phys.: Condens. Matter}
  \textbf{\bibinfo{volume}{19}}, \bibinfo{pages}{116105}
  (\bibinfo{year}{2009}).

\bibitem[{\citenamefont{Szortyka and Barbosa}(2007)}]{Sz07}
\bibinfo{author}{\bibfnamefont{M.~M.} \bibnamefont{Szortyka}} \bibnamefont{and}
  \bibinfo{author}{\bibfnamefont{M.~C.} \bibnamefont{Barbosa}},
  \bibinfo{journal}{Physica A} \textbf{\bibinfo{volume}{380}},
  \bibinfo{pages}{27} (\bibinfo{year}{2007}).

\bibitem[{\citenamefont{Girardi
  et~al.}(2007{\natexlab{a}})\citenamefont{Girardi, Balladares, Henriques, and
  Barbosa}}]{Gi07a}
\bibinfo{author}{\bibfnamefont{M.}~\bibnamefont{Girardi}},
  \bibinfo{author}{\bibfnamefont{A.~L.} \bibnamefont{Balladares}},
  \bibinfo{author}{\bibfnamefont{V.}~\bibnamefont{Henriques}},
  \bibnamefont{and} \bibinfo{author}{\bibfnamefont{M.~C.}
  \bibnamefont{Barbosa}}, \bibinfo{journal}{J. Chem. Phys.}
  \textbf{\bibinfo{volume}{126}}, \bibinfo{pages}{064503}
  (\bibinfo{year}{2007}{\natexlab{a}}).

\bibitem[{\citenamefont{Szortyka et~al.}(2009)\citenamefont{Szortyka, Girardi,
  Henriques, and Barbosa}}]{Sz09}
\bibinfo{author}{\bibfnamefont{M.~M.} \bibnamefont{Szortyka}},
  \bibinfo{author}{\bibfnamefont{M.}~\bibnamefont{Girardi}},
  \bibinfo{author}{\bibfnamefont{V.}~\bibnamefont{Henriques}},
  \bibnamefont{and} \bibinfo{author}{\bibfnamefont{M.~C.}
  \bibnamefont{Barbosa}}, \bibinfo{journal}{J. Chem. Phys.}
  \textbf{\bibinfo{volume}{130}}, \bibinfo{pages}{094504}
  (\bibinfo{year}{2009}).

\bibitem[{\citenamefont{Szortyka et~al.}(2010)\citenamefont{Szortyka, Girardi,
  Henriques, and Barbosa}}]{Sz10a}
\bibinfo{author}{\bibfnamefont{M.~M.} \bibnamefont{Szortyka}},
  \bibinfo{author}{\bibfnamefont{M.}~\bibnamefont{Girardi}},
  \bibinfo{author}{\bibfnamefont{V.}~\bibnamefont{Henriques}},
  \bibnamefont{and} \bibinfo{author}{\bibfnamefont{M.~C.}
  \bibnamefont{Barbosa}}, \bibinfo{journal}{J. Chem. Phys.}
  \textbf{\bibinfo{volume}{132}}, \bibinfo{pages}{134904}
  (\bibinfo{year}{2010}).

\bibitem[{\citenamefont{Girardi
  et~al.}(2007{\natexlab{b}})\citenamefont{Girardi, Szortyka, and
  Barbosa}}]{Gi07}
\bibinfo{author}{\bibfnamefont{M.}~\bibnamefont{Girardi}},
  \bibinfo{author}{\bibfnamefont{M.}~\bibnamefont{Szortyka}}, \bibnamefont{and}
  \bibinfo{author}{\bibfnamefont{M.~C.} \bibnamefont{Barbosa}},
  \bibinfo{journal}{Physica A} \textbf{\bibinfo{volume}{386}},
  \bibinfo{pages}{692} (\bibinfo{year}{2007}{\natexlab{b}}).

\bibitem[{\citenamefont{Buzano et~al.}(2008)\citenamefont{Buzano, De~Stefanis,
  and Pretti}}]{Bu08}
\bibinfo{author}{\bibfnamefont{C.}~\bibnamefont{Buzano}},
  \bibinfo{author}{\bibfnamefont{E.}~\bibnamefont{De~Stefanis}},
  \bibnamefont{and} \bibinfo{author}{\bibfnamefont{M.}~\bibnamefont{Pretti}},
  \bibinfo{journal}{J. Chem. Phys.} \textbf{\bibinfo{volume}{129}},
  \bibinfo{pages}{024506} (\bibinfo{year}{2008}).

\bibitem[{\citenamefont{Binder}(1981)}]{Bi81}
\bibinfo{author}{\bibfnamefont{K.}~\bibnamefont{Binder}}, \bibinfo{journal}{Z.
  Phys. B} \textbf{\bibinfo{volume}{43}}, \bibinfo{pages}{119}
  (\bibinfo{year}{1981}).

\bibitem[{\citenamefont{Landau and Binder}(2003)}]{Landau}
\bibinfo{author}{\bibfnamefont{D.~P.} \bibnamefont{Landau}} \bibnamefont{and}
  \bibinfo{author}{\bibfnamefont{K.}~\bibnamefont{Binder}},
  \emph{\bibinfo{title}{A Guide to Monte Carlo Simulations in Statistical
  Physics}} (\bibinfo{publisher}{Cambridge University Press},
  \bibinfo{address}{New York}, \bibinfo{year}{2003}), \bibinfo{edition}{1st}
  ed.

\bibitem[{\citenamefont{Kolomeisky and Widom}(1999)}]{Ko99}
\bibinfo{author}{\bibfnamefont{A.~B.} \bibnamefont{Kolomeisky}}
  \bibnamefont{and} \bibinfo{author}{\bibfnamefont{B.}~\bibnamefont{Widom}},
  \bibinfo{journal}{Faraday Discussions} \textbf{\bibinfo{volume}{112}},
  \bibinfo{pages}{81} (\bibinfo{year}{1999}).

\bibitem[{\citenamefont{Widom et~al.}(2003)\citenamefont{Widom, Bhimalapuram,
  and Koga}}]{Wi03}
\bibinfo{author}{\bibfnamefont{B.}~\bibnamefont{Widom}},
  \bibinfo{author}{\bibfnamefont{P.}~\bibnamefont{Bhimalapuram}},
  \bibnamefont{and} \bibinfo{author}{\bibfnamefont{K.}~\bibnamefont{Koga}},
  \bibinfo{journal}{Phys. Chem. Chem. Phys.} \textbf{\bibinfo{volume}{5}},
  \bibinfo{pages}{3085} (\bibinfo{year}{2003}).

\bibitem[{\citenamefont{Barbosa and Widom}(2010)}]{Ba10}
\bibinfo{author}{\bibfnamefont{M.~A.} \bibnamefont{Barbosa}} \bibnamefont{and}
  \bibinfo{author}{\bibfnamefont{B.}~\bibnamefont{Widom}}, \bibinfo{journal}{J.
  Chem. Phys.} \textbf{\bibinfo{volume}{132}}, \bibinfo{pages}{214506}
  (\bibinfo{year}{2010}).

\bibitem[{\citenamefont{Kikuchi}(1974)}]{kiku}
\bibinfo{author}{\bibfnamefont{R.}~\bibnamefont{Kikuchi}}, \bibinfo{journal}{J.
  Chem. Phys.} \textbf{\bibinfo{volume}{60}}, \bibinfo{pages}{1071}
  (\bibinfo{year}{1974}).

\end{thebibliography}

\end{document}